\documentclass[onecolumn,12pt,preprintnumbers,eqsecnum,amsmath,amssymb]{revtex4}

\usepackage{epsf}
\usepackage{graphicx}  
\usepackage{dcolumn}   
\usepackage{bm}        

\newcommand{\be}{\begin{equation}}
\newcommand{\en}{\end{equation}}
\newcommand{\ba}{\begin{array}}
\newcommand{\ea}{\end{array}}
\newcommand{\bea}{\begin{eqnarray}}

\newcommand{\eea}{\end{eqnarray}}

\begin{document}

\preprint{KU-TP 001}
\preprint{hep-th/0604088}

\title{ Black Holes in Pure Lovelock Gravities}
\author{Rong-Gen Cai\footnote{E-mail address: cairg@itp.ac.cn}}
\affiliation{Institute of Theoretical Physics, Chinese Academy of
Sciences, P.O. Box 2735, Beijing 100080, China}

 \author{Nobuyoshi Ohta\footnote{E-mail address: ohtan@phys.kindai.ac.jp
 }}
\affiliation{Department of Physics, Kinki University,
Higashi-Osaka, Osaka 577-8502, Japan}


\begin{abstract}
The Lovelock gravity is a fascinating extension of general
relativity, whose  action  consists of the dimensionally extended
Euler densities. Compared to other higher order derivative gravity
theories, the Lovelock gravity is attractive since it has a lot of
remarkable features such as that there are no more than second
order derivatives with respect to metric in its equations of
motion, and that the theory is free of ghost. Recently in the
study of black string and black brane in the Lovelock gravity, a
special class of Lovelock gravity is considered, which is named
pure Lovelock gravity, where only one Euler density term exists.
In this paper we study black hole solutions in the special class
of Lovelock gravity and associated thermodynamic properties.  Some
interesting features are found, which are quite different from the
corresponding ones in general relativity.

\end{abstract}

 \maketitle

\newpage
\section{Introduction}
 Over the past years there has been a lot of
interest in black holes in higher derivative gravity theories. It
is so partly due to the AdS/CFT correspondence, where the higher
derivative terms can be regarded as the corrections of large $N$
expansion in the dual conformal field theory, and partly due to
the brane world scenario, where TeV black holes are expected to be
produced in colliders. Thus it is natural to study the effects of
higher derivative curvature terms (see for example~\cite{Tev} and
references therein). Among the higher derivative gravity theories,
the so-called Lovelock gravity \cite{Love} is rather special. The
Lagrangian  of Lovelock gravity consists of the dimensionally
extended Euler densities
\begin{equation}
\label{1eq1}
{\cal L} = \sum^m_{n=0}c_n {\cal L}_n,
\end{equation}
where $c_n$ are arbitrary constants and ${\cal L}_n$ are the
Euler densities of a $2n$-dimensional manifold
\begin{equation}
{\cal L}_n=\frac{1}{2^n}\delta^{a_1b_1\cdots
a_nb_n}_{c_1d_1\cdots c_nd_n}R^{c_1d_1}_{~~~~ a_1b1}\cdots
R^{c_nd_n}_{~~~~ a_nb_n},
\end{equation}
the generalized delta function is totally antisymmetric in both
sets of indices. ${\cal L}_0$ represents the identity, so the constant
$c_0$ is just the cosmological constant. ${\cal L}_1$ gives us the
usual curvature scalar term, while ${\cal L}_2$ is just the
Gauss-Bonnet term. Usually in order for the Einstein general relativity
to be recovered in the low energy limit, the constant $c_1$ must be
positive. For simplicity one may take $c_1=1$.
Since the action of Lovelock gravity is the
sum of the dimensionally extended Euler densities, it is found
that there are no more than second order derivatives with respect
to metric in its equations of motion. Furthermore, the Lovelock
gravity is shown to be free of ghost when expanded on a flat
space, evading any problems with unitarity~\cite{Des,Zwie}.
It is also known that these terms arise with positive coefficients
as higher order corrections in superstring theories, and their
implications for cosmology have been studied~\cite{MO}.

In the literature, the so-called Gauss-Bonnet gravity, containing first
three terms in (\ref{1eq1}), has been intensively discussed. The
spherically symmetric black hole solutions in the Gauss-Bonnet
gravity have been found in \cite{Des,Whee} and discussed \cite{Myers},
and topological nontrivial black holes have been studied in \cite{Cai1}.
The Gauss-Bonnet black holes in de Sitter space have been discussed
separately in \cite{Cai2}. See also \cite{others} for some other
extensions including perturbative AdS black hole solutions in the gravity
theories with second order curvature corrections. In addition, the references
in \cite{Cai3} have investigated the holographic properties associated
with the Gauss-Bonnet theory.

With many terms, the Lagrangian (\ref{1eq1}) is complicated. But
the static, spherically symmetric black hole solutions can indeed
be found~\cite{Whee}, by solving for a real root of a polynomial
equation for the metric function of the solution. Such black hole
solutions have been generalized to the case with nontrivial
horizon topology in \cite{Cai4}. Since there are $m$
($m=[(d-1)/2]$) ($[N]$ denotes the integral part of the number
$N$) coefficients $c_n$ in (\ref{1eq1}), it is quite difficult to
analyze the black hole solution and to extract physical
information from the solution. In \cite{Ban} a set of special
coefficients has been chosen so that the metric function has a
simple expression. In odd dimensions the action is the
Chern-Simons form for the AdS group and in even dimensions it is
Euler density constructed with the Lorentz part of the AdS
curvature tensor. Thus the $m$ Lovelock coefficients are reduced
to two independent parameters: a cosmological constant and a
gravitational constant. Rewrite the Lagrangian (\ref{1eq1}) in the
form~\cite{Ban}
\begin{equation}
\label{1eq2}
{\cal I}= \kappa \sum^m_{n=0}\alpha_n {\cal I}_n,
\end{equation}
where
\begin{equation}
{\cal I}_n= \int \varepsilon_{a_1\cdots a_d}R^{a_1a_2}\wedge
\cdots \wedge R^{a_{2n-1}a_{2n}}\wedge e^{a_{2n+1}}\wedge \cdots
 e^{a_d}.
\end{equation}
These coefficients $\alpha_n$ are given by
\begin{equation}
\label{1eq4}
\alpha_n = \left \{
\begin{array}{l}
\frac{1}{d-2n}\left (\begin{array}{c}
m-1 \\
n
\end{array}
\right )l^{-d+2n} \ ~~~ {\rm for\ } d= 2m-1 \\
\left ( \begin{array}{c}
m \\
n \end{array}
\right) l^{-d+2n} \ ~~~~~~~~~~~~~~~ {\rm for\ }  d= 2m,
\end{array} \right.
\end{equation}
where $l$ is a length and $\kappa$ in (\ref{1eq2}) is another
parameter. The static, spherically symmetric black hole
solution in the theory (\ref{1eq2}) has a very simple form
\begin{equation}
\label{1eq6}
ds^2= -f(r) dt^2 +f(r)^{-1}dr^2 + r^2 d\Omega_{d-2}^2,
\end{equation}
where
\begin{equation}
\label{1eq7}
f(r) = \left \{
\begin{array}{ll}
1-(2M/r)^{\frac{1}{m-1}} +(r/l)^2 & {\rm for\ } d=2m \\
1-(M+1)^{\frac{1}{m-1}} +(r/l)^2 & {\rm for\ } d=2m-1,
\end{array}
\right.
\end{equation}
where $M$ is an integration constant, interpreted as the mass of
the black hole solution. The nontrivial topological black holes in
this gravity has been studied in \cite{CS}. On the other hand, the
authors of ~\cite{Cris} have chosen a set of coefficients so that
the gravity theory has a unique AdS vacuum with a fixed
cosmological constant and the theory is labelled by an integer $i$
(in \cite{Cris} the integer is denoted by $k$, in this paper,
however, the symbol $k$ will be used for another purpose). In that
case, the black hole solution has also a simple expression. The
coefficients chosen in \cite{Cris} are
\begin{equation}
\label{1eq8}
\alpha_n = \left \{
\begin{array}{ll}
\frac{l^{2(n-i)}}{d-2n} \left(
\begin{array}{c}
i \\
n \end{array}
\right), & n \le i
\\
0,& n>i
\end{array}
\right.
\end{equation}
where the integer $i$ is in the range $1 \le i \le [(d-1)/2]$.
In that theory  the black hole solution has the form (\ref{1eq6}),
but with
\begin{equation}
\label{1eq9}
f(r)= 1+\frac{r^2}{l^2} -\sigma
\left(\frac{C_1}{r^{d-2i-1}}\right)^{1/i},
\end{equation}
where $C_1$ is an integration constant related to the mass of the
black hole solution, and $\sigma =(\pm1)^{i+1}$. We can see
from the solution that when $i=m-1$, the solution (\ref{1eq9})
reduces to (\ref{1eq7}) in even dimensions.

The black strings and black branes are generalized configurations
of black holes, they are some extended objects covered by event
horizon in  transverse directions of these extended objects. These
black configurations play a key role in establishing the AdS/CFT
correspondence.  In the vacuum Einstein gravity, it is easy to
construct black string and black brane solutions by simply adding
some Ricci flat directions to a Schwarzschild black hole solution
or its rotating generalization.  However, it turns out not trivial
to find the black string and black brane solutions in the Lovelock
gravity. It was first noticed in \cite{Bar} that such a simple
method does not work for the Gauss-Bonnet gravity, instead some
numerical approaches have to be adopted~\cite{strings}. More
recently it has been independently realized by Kastor and
Mann~\cite{KM} and Giribet {\it et al.} \cite{Gir} that to
construct some simple black string and black brane solutions in
the Lovelock gravity, the so-called pure Lovelock gravity has to
be invoked, in particular in the case of the asymptotically flat
black string and black brane solutions. Simply speaking, the
action of pure Lovelock gravity is just the one (\ref{1eq1}), but
only one of those coefficients does not vanish.

It is well known that some thermodynamic properties of black
string and black brane solutions can be obtained by studying
thermodynamics of the corresponding black holes, which come from
the dimensional reductions along the isometric directions of black
string and black brane solutions. In this paper we will therefore
study black hole solutions in the pure Lovelock gravity. On the
other hand, due to the characteristic role of black holes in
quantum gravity, studying black hole solution in the pure Lovelock
gravity might be of interest in its own right, for example, in
order to find the difference of black hole solutions  in general
relativity and in pure Lovelock gravity.  We notice that the black
hole solution has been studied in Weyl conformal
gravity~\cite{Weyl}.

In the Lagrangian (\ref{1eq1}), the cosmological constant term
${\cal L}_0$ appears as an independent term.  In this paper,
we will discuss black hole solutions in the theory with only an
Euler density term (\ref{1eq1}) plus the cosmological
constant term. Since the case of $n=1$ is just the Einstein
general relativity, we will mainly discuss the case with $n>1$.

The organization of the paper is as follows. In the next section
we will present the black hole solution in pure Lovelock gravity,
 and study associated thermodynamic properties. In Sec.~IIA, IIB
and IIC, we discuss the cases with vanishing, positive and
negative cosmological constant, respectively. The conclusions and
discussions are given in Sec.~III.

\section{Black Hole Solutions in Pure Lovelock Gravity}

Consider the gravity theory whose Lagrangian consists of the
cosmological constant term ${\cal L}_0$ and the Euler density term
${\cal L}_n$ with $1 \le n\le m$,
\begin{equation}
\label{2eq1}
{\cal L}= c_0 +c_n {\cal L}_n.
\end{equation}
Then the equations of motion are: ${\cal G}_{ab}=0$,
where~\cite{KM}
\begin{equation}
{\cal G}^a_b= c_0\delta^a_b + c_n \delta^{ac_1\cdots c_nd_1\cdots
d_n}_{be_1\cdots e_kf_1\cdots f_n}R^{e_1f_1}_{~~~~c_1d_1}\cdots
R^{e_nf_n}_{~~~~c_nd_n}.
\end{equation}
Assume that the metric is of the form
\begin{equation}
\label{2eq3}
ds^2 =-f(r) dt^2 + f(r)^{-1}dr^2 +r^2 d\Sigma_{d-2}^2,
\end{equation}
where $d\Sigma_{d-2}^2$ is the line element for a
$(d-2)$-dimensional Einstein manifold with constant curvature
scalar $(d-2)(d-3)k$. Here $k$ is a constant, and without loss of
generality, one may take $k=0$ or $\pm 1$. For the theory
(\ref{2eq1}), the solution can be expressed
by~\cite{Cai4,Whee}
\begin{equation}
\label{2eq4}
f(r)=k-r^2F(r),
\end{equation}
where $F(r)$ is determined by the equation
\begin{equation}
\label{2eq5}
\hat{c}_0 +\hat{c}_n F^n(r)=\frac{16\pi G M}{(d-2)\Sigma_{d-2}
r^{d-1}}
\end{equation}
where $G$ is the Newtonian constant in $d$ dimensions,
$\Sigma_{d-2}$ is the volume of the $(d-2)$-dimensional Einstein
constant curvature manifold.  $M$ is an integration constant,
which is in fact the mass of the solution according to the
Hamiltonian method~\cite{Ban,CS}. In addition, we have
\begin{eqnarray}
&& \hat c_0 =\frac{c_0}{(d-1)(d-2)}, \ \ \ \hat c_1=1, \nonumber
\\
&& \hat c_n=c_n\Pi^{2m}_{j=3}(d-j), \ \ \ {\rm for \ } n>1.
\end{eqnarray}
Note that the parameter $c_n$ has the dimension $[{\rm
length}]^{2n-2}$. Since only one parameter $\hat c_n$  appears in
(\ref{2eq5}), except for the cosmological constant $\hat c_0$, we
may normalize the parameter $c_n(>0)$ so that one has $\hat
c_n=\alpha^{2n-2} $ for simplicity, where $\alpha$ is a length
scale. Here we have assumed that $c_n >0$, as in the case of
general relativity and higher derivative terms  in superstring
theories. Furthermore, we set $\hat c_0= -1/l^2$ and then find
that the solution has the form
\begin{equation}
\label{2eq7} F(r) =\left \{
\begin{array}{ll}
\pm \frac{1}{\alpha^{2-2/n}} \left(\frac{16\pi G M}{(d-2)\Sigma_{d-2}
r^{d-1}} +\frac{1}{l^2}\right)^{1/n}  & {\rm for \ } n=\mbox{even}, \\
\frac{ sign(x)}{\alpha^{2-2/n}} \left|\frac{16\pi G
M}{(d-2)\Sigma_{d-2}
r^{d-1}} +\frac{1}{l^2}\right|^{1/n} & {\rm for\ } n=\mbox{odd},
\end{array} \right.
\end{equation}
where  $x \equiv \frac{16\pi G M}{(d-2)\Sigma_{d-2} r^{d-1}}
+\frac{1}{l^2}$.  As expected, when $n=1$, the solution is just
the one describing a Schwarzschild black hole in AdS(dS) space.
When $n <(d-1)/2$, except the singularity at $r=0$, the solution
(\ref{2eq4}) with (\ref{2eq7}) has a potential singularity at
$x=0$.  To see this, let us calculate the Riemann tensor squared
for the metric (\ref{2eq3}):
\begin{equation}
\label{R^2}
 R_{abcd}R^{abcd}=(f''(r))^2
+\frac{2(d-2)}{r^2}(f'(r))^2+\frac{2(d-2)(d-3)}{r^4}(k-f(r))^2,
\end{equation}
where a prime denotes the derivative with respect to $r$.
It is easy to see that the Riemann tensor squared diverges if $x=0$ at some
point $r_x$. Therefore we always consider the region $r>r_x$ in
what follows, if there exists this singularity.

When $n$ is even, the solution becomes $F(r)=\pm
\frac{1}{\alpha^{2-2/n}} \left(
\frac{1}{l^2}\right)^{1/n}$ for $M \to 0$. Therefore in order to have a
well-defined vacuum solution, the cosmological constant $l^2$
must be positive, otherwise the theory is not well-defined.
That is, in this case, the pure Lovelock gravity (\ref{2eq1}) must have
a positive cosmological constant (note that usually the cosmological
constant appears in Lagrangian like ${\cal L}= -2\Lambda +\cdots$).
On the other hand, when $n$ is odd, in order to have a
well-defined vacuum solution $(M \to 0)$, the sign of $x$ must be
the same as the one of $l^2$. In addition, when $l^2=0$, the
solution reduces to
\begin{equation}
\label{2eq8} F(r) = \left \{
\begin{array}{ll}
\pm \frac{1}{\alpha^{2-2/n}} \left(\frac{16\pi G M}{(d-2)\Sigma_{d-2}
r^{d-1}}\right)^{1/n}  & {\rm for \ } n= \mbox{even}, \\
\frac{ sign(M)}{\alpha^{2-2/n}} \left|\frac{16\pi G
M}{(d-2)\Sigma_{d-2}
r^{d-1}}\right|^{1/n} & {\rm for\ } n= \mbox{odd},
\end{array} \right.
\end{equation}
where $M$  must be positive when $n$ is even. In the following
subsections we will discuss the cases $1/l^2=0$,
$1/l^2>0$ and $1/l^2<0$, respectively.

\subsection{The case with a vanishing cosmological constant $1/l^2=0$}

In this subsection, we consider the case without the cosmological
constant, namely the case $1/l^2 =0$. A related solution has been
also considered in~\cite{Char}.   In this case, we see from
(\ref{2eq8}) that the solution is asymptotically flat for $n<
(d-1)/2$. When $n=(d-1)/2$,  the solution exists only in odd
dimensions and describes a topological defect. The metric function
$f(r)$ is constant:
\begin{equation}
\label{2eq9} f(r) =\left \{
\begin{array}{ll}
k \mp \frac{1}{\alpha^{2-2/n}} \left(\frac{16\pi G M}{(d-2)\Sigma_{d-2}
}\right)^{1/n}  & {\rm for \ } n=\mbox{even}, \\
k-\frac{ sign(M)}{\alpha^{2-2/n}} \left|\frac{16\pi G
M}{(d-2)\Sigma_{d-2} }\right|^{1/n} & {\rm for\ } n=\mbox{odd}.
\end{array} \right.
\end{equation}
Note that although the metric function $f$ is a constant in this
case, only when $d=3$, the spacetime is locally flat, and when
$d>3$, some scalar invariants diverge at the origin as can be seen
from (\ref{R^2}). When $k=1$, we see from the solution that the
solid deficit angle is negative in the plus branch in (\ref{2eq9})
for even $n$ and $M<0$ for odd $n$, while the solution just
corresponds to coordinate rescalings when $k=0$ and $-1$, since
there are no fixed periods for coordinates of the Einstein
manifolds in these two cases.

When $n<(d-1)/2$, the solution describes a naked singularity or
black hole,
\begin{equation}
\label{2eq10}
 f(r) =\left \{
\begin{array}{ll}
k \mp \frac{r^2}{\alpha^{2-2/n}} \left(\frac{16\pi G M}{(d-2)\Sigma_{d-2}
r^{d-1}}\right)^{1/n}  & {\rm for \ } n=\mbox{even}, \\
k-\frac{ sign(M)r^2}{\alpha^{2-2/n}} \left|\frac{16\pi G
M}{(d-2)\Sigma_{d-2}
r^{d-1}}\right|^{1/n} & {\rm for\ } n=\mbox{odd}.
\end{array} \right.
\end{equation}
When $k=0$, there is always a naked singularity in the solution.
When $k=-1$, a naked singularity with a cosmological horizon
appears in the plus branch in (\ref{2eq10}) for even $n$ and $M<0$
for odd $n$; the naked singularity has no cosmological horizon for
the minus branch and $M>0$ for odd $n$. When $k=1$, the naked
singularity appears in the plus branch in (\ref{2eq10}) for even
$n$ and $M<0$ for odd $n$. The naked singularity is of little
physical interest. We therefore turn to the black hole solution.
The black hole horizon appears for positive mass $M>0$ in both
cases. Thus we can uniformly rewrite the black hole solution as
\begin{equation}
\label{2eq11}
 f(r)=1-r^2\left(\frac{16\pi G
M\alpha^{2-2n}}{(d-2)\Sigma_{d-2} r^{d-1}}\right)^{1/n}.
\end{equation}
We notice that this solution is just the asymptotically flat limit
discussed in the first reference of \cite{Cris}\footnote{We thank
the referee for pointing  this to us.}. The black hole has a
horizon at $r=r_+$,
\begin{equation}
\label{2eq12}
r_+ = \left( \frac{16\pi G M\alpha^{2-2n}}{(d-2)
\Sigma_{d-2}}\right)^{\frac{1}{d-2n-1}}.
\end{equation}
The black hole has a Hawking temperature, which can be obtained by
continuing the black hole solution (\ref{2eq3}) to its Euclidean
sector, and requiring the absence of conical singularity at the
black hole horizon, which leads to a fixed period of the Euclidean
time, namely the inverse Hawking temperature of the black hole. It
is given by
\begin{equation}
\label{2eq13}
 T=\frac{d-2n-1}{4\pi n}\frac{1}{r_+}.
\end{equation}
The black hole has an associated entropy with the horizon. In
Einstein general relativity, the black hole entropy obeys the
so-called horizon area formula. But in the higher derivative
gravity theories, it is no longer true. The black hole as
a thermodynamic system, must obey the first law of thermodynamics.
Therefore we can use the first law to obtain the black hole
entropy in the pure Lovelock gravity, as we did in the case of
general Lovelock gravity~\cite{Cai4}. According to the first law
\begin{equation}
\label{2eq14}
dM =TdS,
\end{equation}
the black hole entropy can be obtained from the following
integration
\begin{equation}
\label{2eq15}
 S= \int_0^{M} T^{-1}dM=\int_0^{r_+}T^{-1}\left
(\frac{\partial M}{\partial r_+}\right) dr_+.
\end{equation}
Here we have assumed that the black hole entropy vanishes as the
horizon radius goes to zero. Substituting (\ref{2eq12}) and
(\ref{2eq13}) into (\ref{2eq15}), we find
\begin{equation}
\label{2eq16} S=\frac{(d-2)\Sigma_{d-2}}{4G}\frac{n
\alpha^{2n-2}}{d-2n}r_+^{d-2n}.
\end{equation}
We see that the entropy obeys the area formula only when $n=1$,
namely the case of Einstein general relativity. It is easy to show
that this entropy of black hole can also be obtained by using the
entropy formula for black holes in Lovelock gravity~\cite{Jacob}.
Compared to the Schwarzschild black hole, the black hole with
$n>1$ in pure Lovelock gravity has smaller entropy. But like the
Schwarzschild black hole, the black hole in pure Lovelock gravity
has always a negative heat capacity
\begin{equation}
C \equiv \frac{\partial M}{\partial T}=- \frac{(d-2)n
\Sigma_{d-2}\alpha^{2n-2} }{4G}r_+^{d-2n},
\end{equation}
which indicates the thermodynamic instability of the black hole.
Since the black hole solution (\ref{2eq11}) is just the one for
asymptotically flat limit in \cite{Cris}, as a result, these
thermodynamic properties we obtained above are completely the
 same as those found in \cite{Cris}.

\subsection{The case with a positive cosmological constant $l^2>0$}

When $l^2>0$, the solutions are written as
\begin{equation}
\label{2eq18} f(r) = \left \{
\begin{array}{ll}
k\mp \frac{r^2}{\alpha^{2-2/n}} \left(\frac{16\pi G M}{(d-2)\Sigma_{d-2}
r^{d-1}} +\frac{1}{l^2}\right)^{1/n}  & {\rm for \ } n=\mbox{even}, \\
k-\frac{ r^2}{\alpha^{2-2/n}} \left|\frac{16\pi G
M}{(d-2)\Sigma_{d-2}
r^{d-1}} +\frac{1}{l^2}\right|^{1/n} & {\rm for\ } n=\mbox{odd}.
\end{array} \right.
\end{equation}
In this case, this solution is asymptotically dS or AdS. When
$n=(d-1)/2$, the solution reduces to
\begin{equation}
\label{2eq19} f(r) = \left \{
\begin{array}{ll}
k\mp \frac{1}{\alpha^{2-2/n}} \left(\frac{16\pi G M}{(d-2)\Sigma_{d-2}
} +\frac{r^{2n}}{l^2}\right)^{1/n}  & {\rm for \ } n=\mbox{even}, \\
k-\frac{1}{\alpha^{2-2/n}} \left|\frac{16\pi G
M}{(d-2)\Sigma_{d-2} } +\frac{r^{2n}}{l^2}\right|^{1/n} &
{\rm for\ } n=\mbox{odd}.
\end{array} \right.
\end{equation}
Clearly this is the topological defect solution in the pure
Lovelock gravity with a positive cosmological constant
(\ref{2eq1}).  We see that the topological defect solution in the
case $n>1$ is quite different from the case of $n=1$.

When $n<(d-1)/2$, which we discuss in what follows, the solution
(\ref{2eq18}) describes a naked singularity or black hole again.

\begin{figure}[hbt]
\includegraphics[totalheight=1.7in]{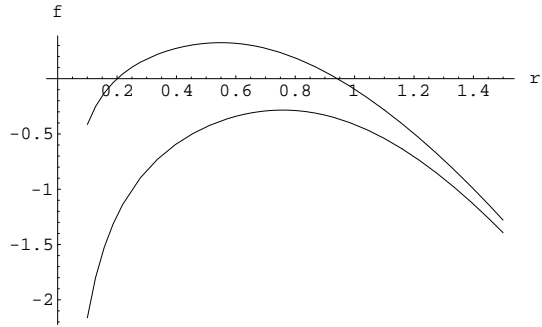}
\caption{ The metric function $f(r)$ (\ref{2eq20}) versus the
radius. The upper curve indicates that both black hole and
cosmological horizons exist, while the lower curve is a naked
singularity solution.} \label{f1}
\end{figure}

(1) When $k=1$, we see that the plus branch in (\ref{2eq18}) for
even $n$  describes a naked singularity. The solution is asymptotically
AdS although the the cosmological constant $l^2$ is positive.
For other cases with $M>0$, the solution (\ref{2eq19}) describes
a black hole. Thus for any $n$ the black hole solution can be written as
\begin{equation}
\label{2eq20}
f(r)= 1-r^2 \left(\frac{16\pi G
M \alpha^{2-2n}}{(d-2)\Sigma_{d-2}
r^{d-1}} + \frac{\alpha^{2-2n}}{l^2}\right)^{1/n},
\end{equation}
In the limit of $r\to \infty $ or $M=0$, the solution
reduces to
\begin{equation}
\label{2eq21}
f(r)= 1- r^2( \alpha^{2-2n}/l^2)^{1/n}.
\end{equation}
This is a dS solution with dS radius
$r_c=(l/\alpha^{1-n})^{1/n}$. Therefore the solution
(\ref{2eq20}) is asymptotically dS and it describes a black hole
in dS space.  This solution is very similar to the
Schwarzschild solution in dS space: in the small $r$ limit, the
first term in the round bracket in (\ref{2eq20}) dominates, while
the second term dominates in the large $r$ limit. We therefore
expect that both black hole and cosmological horizons
appear in a suitable parameter space (see Fig.~\ref{f1}). Both horizons
satisfy the equation $f(r)=0$: the smaller root denotes the black
hole horizon $r_+$ while the larger one $r_c$ corresponds to the cosmological
horizon.

The black hole mass $M$ can be expressed by the black
hole horizon $r_+$ as
\begin{equation}
\label{2eq22}
M= \frac{(d-2)\Sigma_{d-2}r_+^{d-2n-1}}{16\pi G
\alpha^{2-2n}}\left (1-\frac{\alpha^{2-2n}r_+^{2n}}{l^2}\right),
\end{equation}
or in terms of the cosmological horizon $r_c$ as
\begin{equation}
\label{2eq23}
M= \frac{(d-2)\Sigma_{d-2}r_c^{d-2n-1}}{16\pi G
\alpha^{2-2n}}\left (1-\frac{\alpha^{2-2n}r_c^{2n}}{l^2}\right).
\end{equation}
Note that both black hole and cosmological horizons are
always less than $(l/\alpha^{1-n})^{1/n}$ as $M\neq 0$. As in the
asymptotically flat case, the Hawking temperatures associated with
black hole and cosmological horizons can be obtained. They are
\begin{equation}
\label{2eq24}
T_{+,c}=\pm \frac{d-2n-1}{4\pi nr_{+,c}}\left
(1-\frac{d-1}{d-2n-1}\frac{\alpha^{2-2n}r_{+,c}^{2n}}{l^2}\right).
\end{equation}

When the  temperature vanishes, the black hole and cosmological horizons
coincide with each other. In that case, the black hole has horizon radius $r_n$
\begin{equation}
r_n=\left(
\frac{d-2n-1}{d-1}\frac{l^2}{\alpha^{2-2n}}\right)^{1/2n}.
\end{equation}
This is the maximal black hole in the pure Lovelock gravity with
a positive cosmological constant, and it is the counterpart of
the Nariai black hole in this pure Lovelock gravity. When the
mass of the solution is beyond the value~$M_n$
\begin{equation}
\label{2eq26}
M> M_n \equiv \frac{n(d-2)\Sigma_{d-2}r_n^{d-2n-1}}{8(d-1)\pi G
\alpha^{2-2n}},
\end{equation}
the solution (\ref{2eq20}) describes a naked singularity.

As for the entropy associated with horizons, it is easy to check that
both the entropies of black hole and cosmological horizons
still have the form (\ref{2eq16}), and they obey the first law of
thermodynamics
\begin{equation}
dM=T_+dS, \ \ \  -dM=T_c dS_c,
\end{equation}
respectively, where $S_c$ is obtained by replacing $r_+$ in
(\ref{2eq16}) with $r_c$.  This further verifies that (black hole
and cosmological) horizon entropy is a function of horizon
geometry only; the cosmological constant does not appear
explicitly in the expressions of horizon entropy. In addition,
the negative sign in the first law of cosmological horizon is due
to the fact that when $M$ increases the cosmological horizon
shrinks and therefore the entropy $S_c$ decreases~\cite{CMZ}. The
heat capacities  of black hole and cosmological horizons are
\begin{equation}
C_{+,c}
=\left(\frac{\partial M}{\partial T_{+,c}}\right)
= \mp \frac{n^2 \pi (d-2) \Sigma_{d-2}
T_{+,c}r_{+,c}^{d-2n+1}}{(d-2n-1)G\alpha^{2-2n}}
\left(
1+\frac{(d-1)(2n-1)}{d-2n-1}\frac{\alpha^{2-2n}r_{+,c}^{2n}}{l^2}\right)^{-1}.
\end{equation}
While the heat capacity of black hole horizon is always negative,
the heat capacity associated with the cosmological horizon is positive
if one views the energy $E$ of the cosmological horizon as $E=-M$
as in \cite{CMZ}. This indicates that black hole horizon is
thermodynamically unstable while the cosmological horizon is stable as
in the case of Schwarzschild black hole in dS space~\cite{Cai2}.

Note that when $M<0$ in (\ref{2eq20}), the black hole horizon
disappears, instead the solution describes a singularity at $x=0$
covered by a cosmological horizon, which is still determined by
$f(r)=0$.

(2) When $k=0$, it is easy to see from (\ref{2eq18}) that the
solution always describes a naked singularity.

(3) When $k=-1$, it is a naked singularity solution again for the
minus branch in (\ref{2eq18}) for even $n$ and for odd $n$.
However, for the plus branch in the case of even $n$, the solution
becomes
\begin{equation}
\label{2eq29}
f(r)=-1 + \frac{r^2}{\alpha^{2-2/n}}
\left(\frac{16\pi G M}{(d-2)\Sigma_{d-2}
r^{d-1}} +\frac{1}{l^2}\right)^{1/n}.
\end{equation}
This solution is asymptotically AdS although the cosmological
constant in the theory (\ref{2eq1}) is positive. Clearly black
hole horizon will appear in this case.
\begin{figure}[ht]
\includegraphics[totalheight=1.7in]{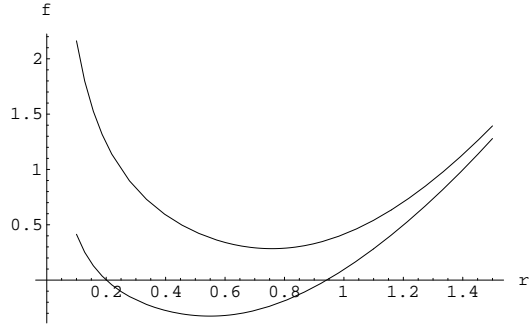}
\caption{ The metric function $f(r)$ (\ref{2eq29}) versus the radius.
The upper curve for a larger $M$ indicates that this solution has a naked
singularity, while there are two black hole horizons for the lower curve.}
\end{figure}

\begin{itemize}

\item  When $M>0$,  the solution has properties as follows. In
Fig.~2 we plot the metric function $f(r)$ versus the radius. From
the figure we can see that in a suitable parameter space, the
solution can have two horizons, one outer horizon and one inner
horizon, while as $M$ becomes large enough, the black hole horizon
disappears and the solution contains a naked singularity.  The
black hole mass $M$ can be expressed in terms of the outer horizon
$r_+$
\begin{equation}
\label{2eq30}
M=\frac{(d-2)\Sigma_{d-2} r_+^{d-1}}{16\pi G}
\left(\frac{\alpha^{2n-2}}{r_+^{2n}}-\frac{1}{l^2}\right ).
\end{equation}
The Hawking temperature of the black hole is
\begin{equation}
\label{2eq31}
T= \frac{d-2n-1}{4\pi n r_+} \left(-1
+\frac{d-1}{d-2n-1}\frac{\alpha^{2-2n}r_+^{2n}}{l^2}\right).
\end{equation}

When the temperature vanishes, the black hole has a horizon
$r_{min}$
\begin{equation}
\label{2eq32} r_{min} \equiv
\left(\frac{d-2n-1}{d-1}\frac{l^2}{\alpha^{2-2n}}\right)^{1/2n}.
\end{equation}

\item When $M=0$, the solution (\ref{2eq29}) still has a black
hole (massless black hole) horizon
\begin{equation}
\label{2eq33}
r_{max} = (l/\alpha^{1-n})^{1/n}.
\end{equation}
Therefore we conclude that the horizon (\ref{2eq32}) is the
minimal one, while the horizon (\ref{2eq33}) is the maximal one if
$M>0$. Here let us mention that when $M=0$, the solution
(\ref{2eq29}) has only a horizon $r_{max}$; when $M$ increases,
two horizons appears; and when $M$ reaches
\begin{equation}
M_{max}=\frac{(d-2)n\Sigma_{d-2}
r_{min}^{d-2n-1}\alpha^{2n-2}}{8(d-1)\pi G},
\end{equation}
two horizons coincide with each other, beyond which the solution
gets a naked singularity. Using the first law of black hole
thermodynamics, $dM=TdS$, we obtain the entropy of the black hole
\begin{equation}
S=C_0 -\frac{(d-2)
\Sigma_{d-2}}{4G}\frac{n}{d-2n}\alpha^{2n-2}r_+^{d-2n},
\end{equation}
where $C_0$ is an integration constant. If one takes $C_0$ to be
zero, one is led to a negative entropy! In fact, the minus sign
arises due to the fact that when the mass $M$ increases, the
horizon radius decreases. This implies that this black hole has a
negative heat capacity; it is thermodynamically unstable.

\item When $M<0$, we can rewrite the solution (\ref{2eq29}) as
\begin{equation}
\label{in}
f (r) =-1 + \frac{r^2}{\alpha^{2-2/n}}
\left(-\frac{16\pi G M'}{(d-2)\Sigma_{d-2}
r^{d-1}} +\frac{1}{l^2}\right)^{1/n},
\end{equation}
where $M'(=-M)>0$ is used. In this case, we have only one black
hole horizon $r_+$, which is always larger than $(l/\alpha
^{1-n})^{1/n}$ for $M'>0$. Namely now the massless black hole
(\ref{2eq33}) becomes a minimal one. For this black hole, a new
singularity appears at $x=0$, that is, $r_x^{d-1}=16\pi GM'
l^2/(d-2)\Sigma_{d-2}$. Note that the singularity is always
covered by the black hole horizon $r_+$. For this black hole we
have Hawking temperature
\begin{equation}
\label{in1}
T= \frac{d-2n-1}{4\pi n r_+}\left
(\frac{d-1}{d-2n-1}\frac{r_+^{2n}}{l^2\alpha^{2n-2}}-1\right),
\end{equation}
and a positive entropy with
\begin{equation}
\label{in2}
S=\frac{(d-2)
\Sigma_{d-2}}{4G}\frac{n}{d-2n}\alpha^{2n-2}r_+^{d-2n}.
\end{equation}
We see from (\ref{in1}) that the Hawking temperature increase when
$r_+$ grows. Therefore this black hole is thermodynamically stable
and has a positive heat capacity.
\end{itemize}

{}From the above analysis, we can see that if one takes the solution
as the form (\ref{in}) with any sign of $M'$, everything goes well
and nothing strange appears: When $M'>0$, the solution
has only one horizon larger than $(l/\alpha ^{1-n})^{1/n}$, the
temperature and entropy are given by (\ref{in1}) and (\ref{in2}),
respectively; when $M'=0$, this is just the massless black hole
with horizon radius $(l/\alpha ^{1-n})^{1/n}$; when $M'<0$, the
black hole solution will have two horizons, the two horizons
coincide with each other when the temperature (\ref{in1})
vanishes. In that case, the black hole has a minimal horizon
(\ref{2eq32}), its mass is negative
$$
M_{min}=-\frac{(d-2)n\Sigma_{d-2}
r_{min}^{d-2n-1}\alpha^{2n-2}}{8(d-1)\pi G},
$$
beyond which the solution describes a naked singularity. This is
nothing but the counterpart of the negative mass hyperbolic black
holes in pure Lovelock gravity (for negative mass hyperbolic black
holes in general relativity see, for example, some references in
\cite{Cai1,others}).

\subsection{The case with a negative cosmological constant $l^2<0$}

Now we turn to the case with a negative cosmological constant
$l^2<0$, namely, $\hat c_0 >0$. In this case, we can see from
(\ref{2eq5}) that when $n$ is even, there is no physical solution
for $F(r)$ unless $\hat c_n <0$, even for the case with $M>0$ in
(\ref{2eq7}), because in the latter case, there is no well-defined
vacuum solution as mentioned above. On the other hand, when $n$ is
odd, the sign of $x$ is also negative (to have a well-defined
vacuum solution again). Combining (\ref{2eq4}) and (\ref{2eq7}),
the solution can be written down as
\begin{equation}
\label{2eq37}
f(r)=k +\frac{ r^2}{\alpha^{2-2/n}}
\left(-\frac{16\pi G M}{(d-2)\Sigma_{d-2}
r^{d-1}} +\frac{1}{|l|^2}\right)^{1/n},
\end{equation}
where $M>0$.
This solution is asymptotically AdS.  It is easy to see that when
$k=1$ or $k=0$, the solution describes a naked singularity, while
$k=-1$, the solution becomes  completely the same as (\ref{in}),
and it becomes the one (\ref{2eq29}) if $M<0$. Namely in the case
of $k=-1$,  a black hole with a negative constant curvature
horizon appears in a suitable parameter space. We will not repeat
the analysis here.

\section{Conclusion and Discussion}

We studied black hole solutions in the pure Lovelock
gravity with a cosmological constant. The Lagrangian
of the pure Lovelock gravity is an Euler density for a certain
spacetime dimension. Such a theory naturally arises in the
construction of black string and black brane solutions in a general
Lovelock gravity~\cite{KM,Gir}. In the case without the
cosmological constant, the solution we found is either a
topological defect solution (\ref{2eq9}), or a black hole
solution (\ref{2eq11}), otherwise it describes a naked
singularity. The black hole thermodynamics was analyzed, and we found that
it has similar properties to a Schwarzschild black hole. When the
cosmological constant is positive, we found black hole solution
(\ref{2eq20}), which is asymptotically dS. The black hole solution
again has similar properties to a Schwarzschild black hole in dS space.
Interestingly enough, in this case, we also found an
asymptotically AdS black hole solution (\ref{2eq29}), which has
a negative constant curvature horizon. When the cosmological
constant is negative, we have not found any solution of physical
interest if the number $n$ is even. However, when $n$ is
odd, we found again the asymptotically AdS black hole solution
with a negative constant curvature horizon.

It is well known that in general relativity, black holes in AdS
space can have positive, zero or negative constant curvature
horizons, namely the cases of $k=1$, $0$ and $-1$. In the pure
Lovelock gravity, however, we have seen that only $k=-1$ black
holes are allowed to appear in the asymptotically AdS space.

Finally we mention that a Maxwell field can be added to
the Lagrangian (\ref{2eq1}). In
this case, a static, spherically symmetric solution (\ref{2eq3})
can be determined by solving the following equation~\cite{Wil,Cai4}
\begin{equation}
\label{3eq1}
\hat c_0 +\hat c_n F^n(r) =\frac{16\pi G M}{(d-2)\Sigma_{d-2}
r^{d-1}}-\frac{q^2}{r^{2(d-2)}}.
\end{equation}
where $q$ is another integration constant, which is related to the
electric charge of the solution. As the case without the charge,
we can also discuss the causal structure of the charged solution
and associated thermodynamic properties.

\section*{Acknowledgments}
This work was finished during RGC's visit to the Department of
Physics at Kinki University through a JSPS invitation fellowship,
the warm hospitality extended to him is appreciated. This work is
supported by grants from NSFC, China (No. 10325525 and No.
90403029), and a grant from the Chinese Academy of Sciences. NO
was supported in part by the Grant-in-Aid for Scientific Research
Fund of the JSPS No. 16540250.


\end{document}